\documentclass[universe,article,accept,pdftex,moreauthors]{Definitions/mdpi} 

\firstpage{1} 
\pubvolume{1}
\issuenum{1}
\articlenumber{0}
\pubyear{2025}
\copyrightyear{2025}
\externaleditor{~}
\datereceived{27 November 2024} 
\daterevised{21 January 2025} 
\dateaccepted{23 January 2025} 
\datepublished{ } 
\hreflink{https://doi.org/} 

\Title{Prediction of Individual Halo Concentrations Across Cosmic Time Using Neural Networks} 

\TitleCitation{Prediction of Individual Halo Concentrations Across Cosmic Time Using Neural Networks} 

\Author{Tianchi Zhang 
 $^{1,}$*\orcidA{}, Tianxiang Mao $^{2}$, Wenxiao Xu
$^{3}$\orcidC{} and Guan Li $^{4,5}$\orcidB{}}

\AuthorNames{Tianchi Zhang, Tianxiang Mao, WenXiao Xu and Guan Li}

\AuthorCitation{Zhang, T.; 
 Mao, T.; Xu, W.; Li, G.}

\address{$^{1}$ \quad Beijing Planetarium, Beijing Academy of Science and Technology, Beijing 100044, China\\
$^{2}$ \quad Key Laboratory for Computational Astrophysics, National Astronomical Observatories, Chinese Academy of Sciences, Beijing 100012, China; maotianxiang@bao.ac.cn \hl{} 
\\
$^{3}$ \quad School of Science, Chongqing University of Posts and Telecommunications, Chongqing 400065, China; xuwx@cqupt.edu.cn 
\\
$^{4}$ \quad Computer Network Information Center, Chinese Academy of Sciences, Beijing 100083, China; liguan@cnic.cn
\\
$^{5}$ \quad University of Chinese Academy of Sciences, Beijing 101408, China
}

\corres{Correspondence: tczhang@bjp.org.cn}

\abstract{The concentration of dark matter haloes is closely linked to their mass accretion history. We utilize the halo mass accretion histories from large cosmological $N$-body simulations as inputs for our neural networks, which we train to predict the concentration of individual haloes at a given redshift. The trained model performs effectively in other cosmological simulations, achieving the root mean square error between the actual and predicted concentrations that significantly lower than that of the model by Zhao et al. and Giocoli et al. at any redshift. This model serves as a valuable tool for rapidly predicting halo concentrations at specified redshifts in large cosmological simulations.}


\begin{document}

\section{Introduction}\label{sec_intro}

{The $\Lambda$CDM model describes the geometry of the universe, and explains how the cosmic web, composed of clusters, filaments, and voids, originated from the Big Bang. This provides the theoretical foundation for cosmological simulations. With the rapid development of supercomputing power and parallel algorithms, large cosmological $N$-body simulations have been made possible, leading to significant advancements in our understanding of the large-scale structures and inner structure of dark matter haloes (see 
\citep{Frenk2012,Angulo2022}, for a brief review)}.

{Haloes form through the collapse of density fluctuations in the dark matter field. At high redshifts, haloes are more compact and exhibit higher concentrations due to the elevated background density of the universe during these epochs. As the universe expands, the growth of haloes slows because of the declining density, leading to less efficient accretion and a decrease in halo concentrations with redshift (see 
\citep{Okoli2017} for a brief review). Recent studies have highlighted the critical role of halo concentration in galaxy formation and evolution. \citet{Jiang2019} employed cosmological hydrodynamic simulations to investigate the relationship between halo properties and galaxy size, their findings indicate that galaxy size is weakly correlated with the angular momentum of the halo~\citep{Mo1998,Yang2023}, it is strongly correlated with halo concentration. Observationally, \citet{Runge2022} studied the fossil group NGC 1600 and revealed that its host halo possesses an exceptionally high concentration. Furthermore, analyses of the gravitational lens system J0946 + 1006 have revealed overconcentrated subhaloes that significantly deviate from predictions of the $\Lambda$CDM model~\citep{Minor2021,enzi2024}.}

Halo concentration is closely related to mass, redshift, and cosmological parameters~\citep{Maccio2008, Ludlow2014}. The relationship between halo mass and halo concentration (hereafter the $c-M$ relation) has been extensively discussed and analyzed by various authors over the past twenty years. A consistent conclusion has been emerged: low-mass haloes exhibit higher concentrations, while high-mass haloes display lower concentrations (see 
 e.g., \citep{Neto2007, Gao2008, Klypin2011, Bhattacharya2013, Dutton2014, Child2018, Diemer2019,Ishiyama2021}, etc.).

The evolution of halo concentration is closely linked to the halo mass accretion history, as discussed in previous studies. \citet{Bullock2001} systematically analyzed halo concentration by relating it to the epoch of initial halo collapse, which determines the initial inner halo density. \citet{Wechsler2002} identified a general framework for describing mass accretion history, establishing a strong correlation between halo concentration and the characteristic formation epoch ($1/(1+z)$). \citet{Zhao2003} found that the mass assembly history of haloes can be roughly divided into an early rapid accretion phase, which establishes the potential well, and a later slow accretion phase, which adds mass without significantly altering the potential well. \citet{Zhao2009} and \citet{Giocoli2012} developed a universal model for the concentration evolution history of dark matter haloes. Their results indicate that concentration evolves with redshift in a more complex manner than suggested by \citet{Wechsler2002}. 

In recent years, the application of machine learning in astronomy has seen significant growth (see 
 \citep{Fluke2020, Sen2022} for a recent review). Machine learning allows us to model complex physical processes, especially nonlinear problems, through simpler frameworks that are valuable in computational cosmology. \citet{Aragon2019} used convolutional neural networks to segment cosmic filaments and walls in the large-scale structure of the universe. \citet{Sun2019} employed the mean-shift algorithm to identify haloes and subhaloes in numerical simulations. \citet{Wadekar2021} investigated the connection between haloes and galaxies using IllustrisTNG simulations. \citet{Mao2021} applied deep learning methods to reconstruct baryonic acoustic oscillation signals, improving initial conditions for cosmological simulations. {\citet{Maltz2024} employed an extremely randomized trees machine learning approach to model the relationship between galaxies and their subhaloes across a wide range of environments in the First, Light and Reionisation Epoch Simulations.}

In this work, we use neural network to predict individual halo concentrations at different redshifts in $N$-body cosmological simulations, and compare the results with other models. This paper is organized as follows: In Section \ref{sec_method}, we detail our numerical simulations, halo sample, and introduce our neural network model. In Section \ref{sec_res}, we present the halo concentrations predicted by the neural network model, and compare them with other models. Finally, we summarize our conclusions {and discussions} in Section \ref{sec_con}.
\section{Simulations and Neural Network Model} \label{sec_method}
\subsection{{Simulations}}
Simulations were run using the Tree-PM code \textsc{Gadget-2}~\citep{springel2005}, with the following cosmological parameters: $\Omega_{\rm m}=0.3, \Omega_\Lambda=0.7, \Omega_{\rm b}=0.04, \sigma_8=0.9, {\rm h}=0.7$, and $n_{\rm s}=0.96$. Initial conditions were set at $z=127$ and generated with the \textsc{N-genic} code based on the linear matter power spectrum from \citet{Eisenstein1998}. Prior to this, we applied the capacity-constrained Voronoi tessellation(CCVT) method~\citep{Liao2018,Zhang2021} to create a uniform and isotropic particle distribution. 

We performed two cosmological simulations: SimA, used to train our neural network model, and SimB, used to test the robustness of the model. SimA consists of $1024^3$ dark matter particles in a periodic box with length of $200 \; h^{-1}{\rm Mpc} $ on each side, yielding a mass resolution of $6.20 \times 10^8 \; h^{-1}{\rm M_\odot}$. SimB has the same box size and cosmological parameters as SimA, but contains $512^3$ dark matter particles, evolving from a different realization, with a mass resolution of $4.96 \times 10^9 \; h^{-1}{\rm M_\odot}$. The force softening length in all simulations are set to $1/50$ mean interparticle separation, aligning with the roughly optimal gravitational softening length for haloes studied in this work, as indicated by \citet{Zhang2019}. To accurately track the accretion history of haloes, we recorded 135 snapshots between $z=35$ and $z=0$, with an approximate universe age interval of 0.1 Gyr between adjacent snapshots.

Haloes in all simulation are identified using the friends-of-friends algorithm with a linking length of 0.2 times the mean interparticle separation~\citep{Davis1985}. We identify subhaloes and construct the main branch of halo merger trees using the \textsc{HBT+} code~\citep{Han2018}. 

\subsection{{Halo Definition, Concentration, and Datasets}}

Haloes are defined as a spherical region with an average density equal to $\Delta_{\rm vir}$ times the mean cosmic density~\citep{Bryan1998},
\begin{equation} \label{eq_vir} \Delta_\mathrm{vir} = 18\pi^2 + 82x - 39x^2, 
\end{equation}
with $x = \Omega_{\rm m}(z) - 1$, a value related to the redshift $z$, and
\begin{equation} \Omega_{\rm m}(z) = \frac{\Omega_{\rm m}(1+z)^3}{\Omega_{\rm m}(1+z)^3 + (1 - \Omega_{\rm m} - \Omega_\Lambda)(1 + z)^2 + \Omega_\Lambda}. 
\end{equation}

The value 
 of $\Delta_{\rm vir}$ varies with redshift, from approximately 180 at high redshift to around 340 at $z=0$. 
  We also define $M_{\rm vir}$, $r_{\rm vir}$, and $N_{\rm vir}$ to indicate the mass, radius, and number of particles in the halo, respectively. {The reason why we use this definition of halo is to facilitate a comparison with the work of \citet{Zhao2009} and \citet{Giocoli2012} (for more details, refer to the third paragraph of Section \ref{sec_res})}.

Halo density profile is commonly described by the Navarro--Frank--White (NFW) profile~\citep{Navarro1996,Navarro1997}, which depends on the radial distance $r$ as follows:
\begin{equation}\label{eq_NFW}
\rho(r) = \frac{4\rho_{\rm s}}{\left(\frac{r}{r_{\rm s}}\right)\left(1 + \frac{r}{r_{\rm s}}\right)^2},
\end{equation}
where $r_{\rm s}$ is the scale radius that divides the halo into inner and outer regions. The logarithmic density slope is $-1$ in the innermost region and $-3$ in the outer region of the halo. The parameter $\rho_{\rm s}$ represents the density at $r = r_{\rm s}$. The halo concentration parameter $c$ can be determined by fitting the NFW density profile, and is defined as
\begin{equation}\label{eq_c} c = \frac{r_{\rm vir}}{r_{\rm s}}. 
\end{equation}

In SimA, to ensure sufficient resolution of haloes across different redshifts, we trace the $z=0$ haloes with $N_{\rm vir} > 7000$ along the main branch until the $N_{\rm vir}$ of the main progenitor drops below 32. Approximately 7000 haloes are identified, allowing us to trace 100\% of the mass accretion history (MAH) up to $z=4.6$. In our study, we use Equation~(\ref{eq_NFW}) to fit the NFW profile using the least squares method for each MAH between $z=2$ and $z=0$ to determine $r_{\rm s}$. To ensure fitting accuracy, haloes must contain more than 500 particles, and we utilize 20 equally spaced logarithmic bins between $0.05r_{\rm vir}$ and $r_{\rm vir}$. SimB follows the same procedure, but since the number of haloes with $N_{\rm vir} > 7000$ at $z=0$ is too small, so we track haloes with $N_{\rm vir} > 2000$, resulting in 1480 haloes that meet this criterion. We also calculate halo concentrations from $z=2$ to $z=0$ using Equation~(\ref{eq_c}). Note that all haloes of MAH are shuffled to avoid sorting by halo mass or other properties.

\subsection{Neural Network Model}\label{sec_method_NN}

\begin{figure}[H]
\includegraphics[width=0.8\linewidth]{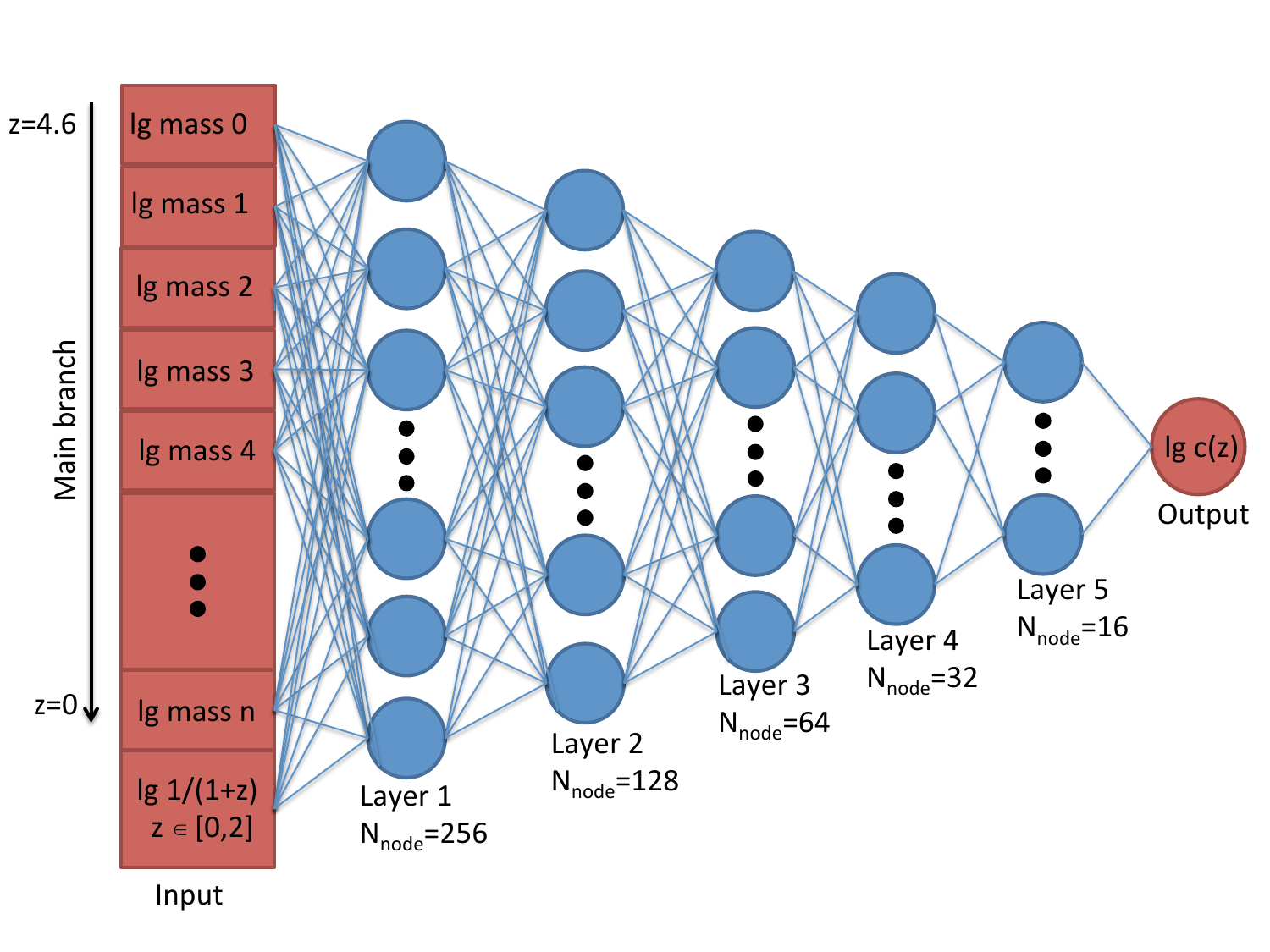} 
\caption{The schematic of our neural network is presented here. Red squares represent the input layer neurons, which take the MAH and the desired $z$ as input parameters. The blue circles indicate the neurons in the hidden layers; there are five layers in total, with the number of neurons in each layer denoted by $\rm N_{ node}$. The red circle represents the single neuron in the output layer, which outputs the halo concentration at $z$.}\label{fig1}
\end{figure}

We employ an optimal framework for the rapid implementation of a neural network using the Python package PyTorch~\citep{pytorch} to predict halo concentrations in this work. The structure of the neural network is illustrated in Figure~\ref{fig1}, which includes the input layer, hidden layer, output layer, and the number of neurons in each layer.

There are 124 neurons in the input layer, which include the halo mass ($\log M_{\rm vir}$) corresponding to 123 snapshots along the MAH and one predicted redshift $z$ (with $z \in [0, 2]$, corresponding to 103 snapshots. Here, we convert $z$ to $\log(1/(1+z))$ to avoid $\log 0$ when $z=0$). In total, the SimA dataset comprises 721,000 datasets, derived from 7000 MAHs multiplied by 103 redshift values. The first 618,000 datasets are used as the TrainDataset to build the model, while the remaining 103,000 serve as the ValidationDataset for multiple iterations to adjust the model's parameters. The 1480 MAHs from SimB consist of a total of 152,440 TestDataset, which are used only once to verify the model's \mbox{generalization performance}. 

We normalize the input parameters using the formula $y=(x-\mu)/\sigma$, where $x$ is the value of each input neuron, and $\mu$ and $\sigma$ represent the mean and standard deviation of the corresponding neurons, respectively. The normalized input parameters $y$ pass through five hidden layers, with the number of nodes in each layer being 256, 128, 64, 32, and 16, respectively. We apply the ReLU activation function ($f(y)=max(0,y)$), which enhances the nonlinearity of the neural network~\citep{Nair2010}. The mean squared error loss function~\citep{Rumelhart1986} is used to measure the error between the actual and predicted values,
\begin{equation}\label{eq_loss}
L=\sum_{\rm i=0}^{\rm BatchSize} (\log c_{\rm i,sim}-\log c_{\rm i,pred})^2,
\end{equation}
where $c_{\rm i,sim}$ is the actual concentration of $i$-th TrainDataset and $c_{\rm i,pred}$ is the concentration predicted by the neural network model, The BatchSize is set to 256 to improve the training speed and enhance randomness in the training process. The weight of each node is updated using back propagation, and the Adam optimizer~\citep{kingma2014} is employed for faster convergence to find the optimal solution after each step within each batch. The output layer exports the final predicted halo concentration at $z$. The neural network model is trained 100 times with a learning rate of 0.001. After each step, we use the ValidationDataset to calculate the root mean square error (RMSE, $\sqrt{L}$) until the value converges, and we save this network as our model for the next section.

\section{Results}\label{sec_res}

\begin{figure}[H]
\includegraphics[width=0.8\linewidth]{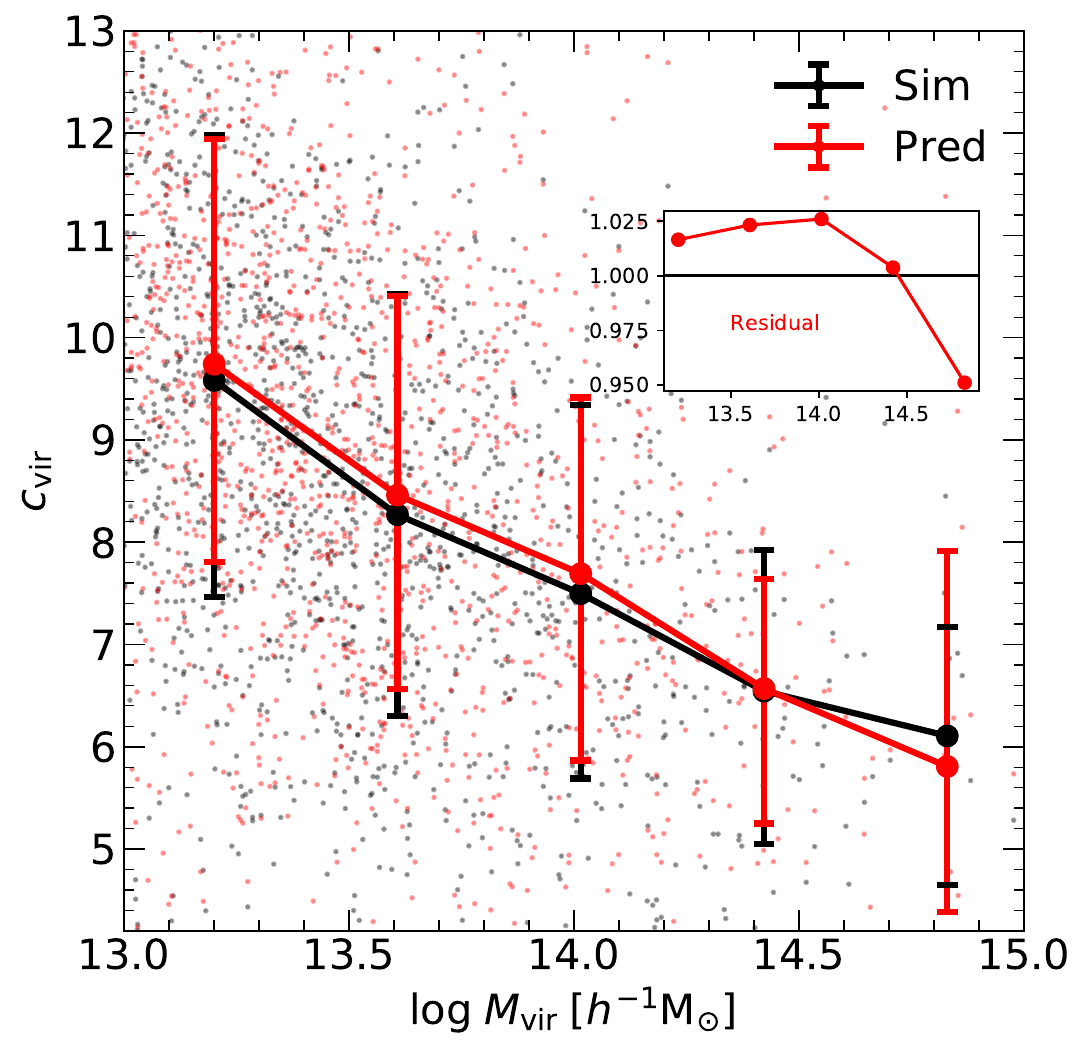} 
\caption{The halo 
 concentration--mass relation at $z=0$ is derived from fitting the NFW profile (black) and utilizing a neural network model (red). The scatter points in the background indicate individual haloes, whereas the larger points connected by lines represent the median values. Error bars denote the 16th and 84th percentiles. The small panel in the upper right displays the residuals between the predicted results of our model and the simulation results.}\label{fig2}
\end{figure}

To assess potential issues with our neural network model, we examined the well-known halo $c-M$ relation to determine whether the model can reproduce universal results: as halo mass increases, halo concentration decreases (see, 
 e.g., \citep{Neto2007, Gao2008, Bhattacharya2013, Dutton2014}, etc.). In Figure~\ref{fig2}, we plot the $c-M$ relation at $z=0$, comparing the fitted NFW profile of haloes in the simulation (black curve) with predictions from our model (red curve). The neural network model closely reproduces the simulation results, supporting the aforementioned universal conclusion. In the residual panel, the median error of our model relative to the median error from the simulation is mostly less than 2.5\%. Only at the massive end does the error increase to about 5\%, likely due to the smaller number of halo samples in this mass range.

\begin{figure}[H]
\includegraphics[width=0.9\linewidth]{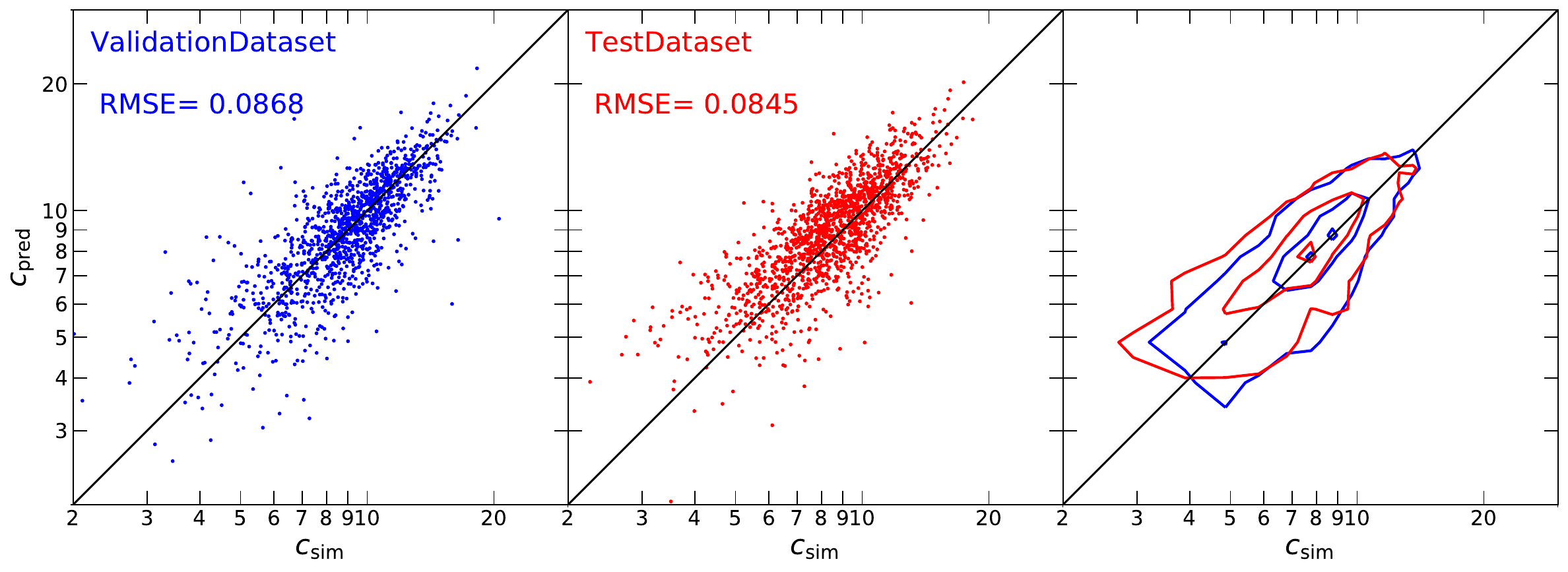} 
\caption{A comparison 
 of the measured concentrations with those predicted by our model in SimA and SimB at $z=0$. Blue points represent the ValidationDataset from SimA, while red points denote the TestDataset from SimB. The contours enclose 10\%, 50\%, and 90\% of the haloes, providing a visual representation of concentration distribution. The solid black lines indicate the $c_{\rm sim}$=$c_{\rm pred}$ in each panel, highlighting the accuracy of the model's predictions. {These highlight the neural network model's ability to generalize effectively, allowing it to predict halo concentrations accurately across different simulations.}}\label{fig3}
\end{figure}

While the statistical performance of halo concentrations obtained from the our model appears promising, we also sought to evaluate its accuracy on an individual halo. In Figure~\ref{fig3}, we plot the halo concentrations measured from simulations (ValidationDataset in SimA and TestDataset in SimB) against those predicted by the our model at $z=0$. The left panel displays a scatter plot for the ValidationDataset (blue), featuring RMSE of 0.0868. The points are closely distributed along the diagonal $c_{\rm sim}$=$c_{\rm pred}$, indicating high predictive accuracy for this dataset. In the middle panel, the model similarly demonstrates strong performance for the TestDataset, achieving an RMSE of 0.0845. The true concentration values and the model's predictions show comparable accuracy for both datasets, with the blue and red points aligning closely along the $c_{\rm sim}$=$c_{\rm pred}$ line. The right panel further illustrates robust agreement between the two contours. These findings emphasize the neural network model's strong generalization capabilities, enabling it to accurately predict halo concentrations not only within the same simulation but also across different simulations. We also tested our model with varying initial condition realizations, and obtained \mbox{similarly predictions}.

\begin{figure}[H]
\includegraphics[width=0.75\linewidth]{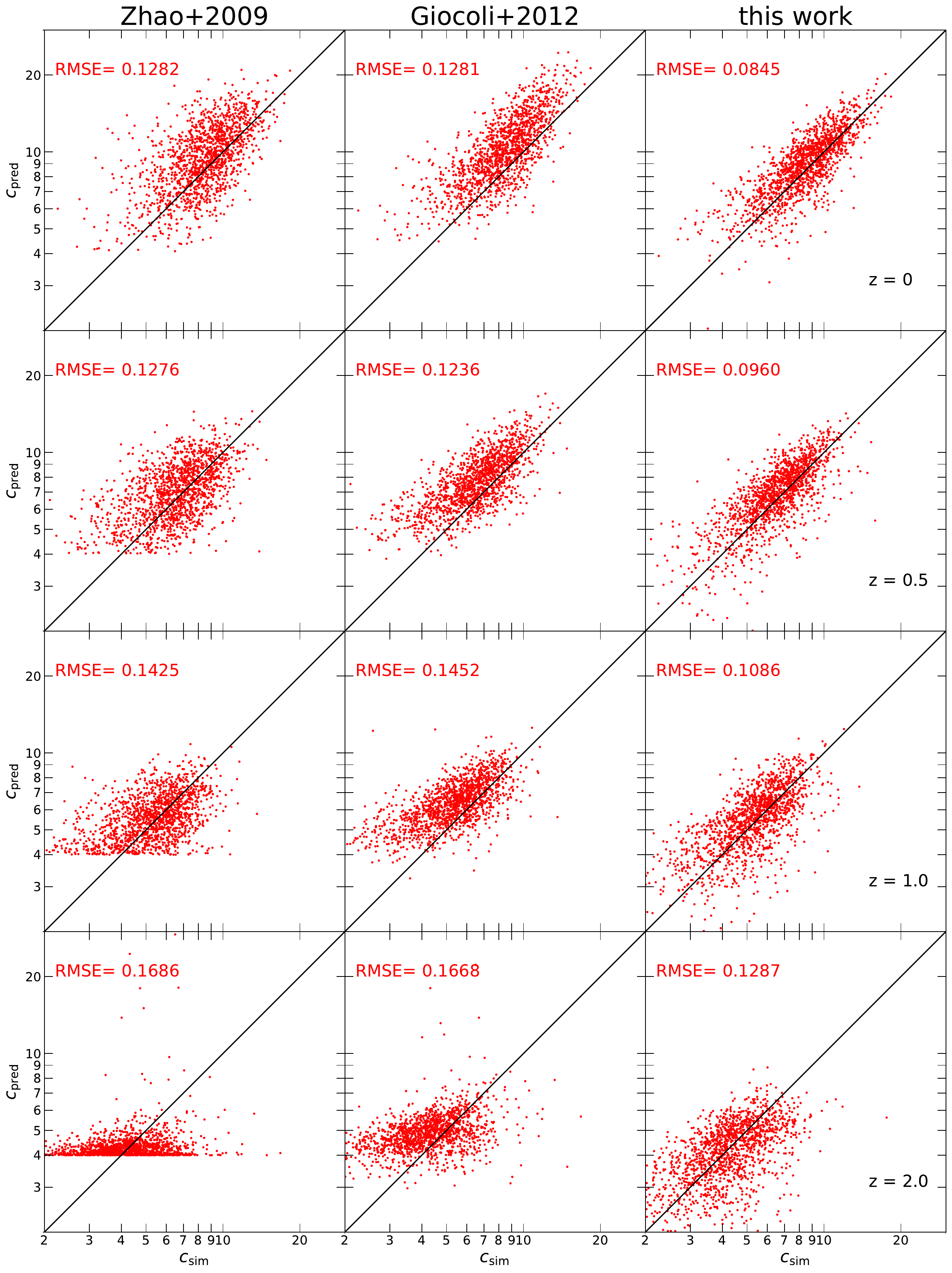} 
\caption{Similar to 
 Figure~\ref{fig3}, except that concentrations predicted by models by \citet{Zhao2009}(denoted as "Zhao+2009", left column), \citet{Giocoli2012}(denoted as "Giocoli+2012", middle column), and neural network in TestDataset at {different redshifts. This demonstrates the model's robust predictive ability across different redshifts.}}\label{fig4}
\end{figure}

For comparison with other works, we derive concentrations from halo mass accretion history using the models of \citet{Zhao2009} and \citet{Giocoli2012}.
In the Zhao et al. model, halo concentrations may be related to the universe is age when the main progenitor first reaches 4\% of its current mass,
\begin{equation}\label{eq_Zhao} c = 4 \left\lbrace1 + \left(\frac{t}{3.75 t_{0.04}}\right)^{8.4}\right\rbrace^{1/8}, 
\end{equation}
where $t_{0.04}$ is the universe is age at that time. The model proposed by Giocoli et al. includes one additional parameter compared to the Zhao et al. model,
\begin{equation} \label{eq_Giocoli} \log c = \log 0.45 \left\lbrace4.23 + \left(\frac{t}{t_{0.04}}\right)^{1.15} + \left(\frac{t}{t_{0.5}}\right)^{2.3} \right\rbrace, 
\end{equation}
where $t_{0.5}$ is the age of the universe when the main progenitor has accreted half of its current mass. The comparison results are shown in Figure~\ref{fig4}. {We first present the results of $z=0$ in the first row.} The left panel presents results from \citet{Zhao2009}, with RMSE of 0.1282. The scatter points are distributed around the diagonal but exhibit significant spread. The middle panel shows results from \citet{Giocoli2012}, yielding RMSE of 0.1281, which displays a similar distribution to \citet{Zhao2009}. In contrast, the right panel depicts results from this work, with RMSE of 0.0845. Here, the scatter points are more tightly clustered around the diagonal, indicating that the neural network model outperforms the previous methods in predicting halo concentrations at $z=0$, while the predictive performance of the other models is comparable, both show slight deviations from the $c_{\rm sim}$=$c_{\rm pred}$ line. In stark contrast, the points predicted by our model closely align with the $c_{\rm sim}$=$c_{\rm pred}$ line across both low and high concentrations, demonstrating the strong predictive power and robustness of our model. {The second, third and fourth rows present the results of $z = 0.5$, 1 and 2. We found that the predictive ability of the neural network model at these redshifts far exceeds that of the other two models. Specifically, at $z=2$ the \citet{Zhao2009} model initializes the concentration of most haloes to 4, leading to unreliable predictions. Meanwhile, the \citet{Giocoli2012} model tends to overestimate the concentration of haloes. In contrast, the scatter in our model remains reasonably distributed around the diagonal, indicating more reliable predictions. What makes our model more effective than the other two? This likely stems from the inclusion of more variables, which provides a richer input for the halo mass accretion history. In contrast, the \citet{Zhao2009} model includes only one variable, $t_{0.04}$, while the \citet{Giocoli2012} model uses two variables, $t_{0.04}$ and $t_{0.5}$. A more detailed mass accretion history leads to more accurate prediction of halo concentration.}

\begin{figure}[H]
\includegraphics[width=0.8\linewidth]{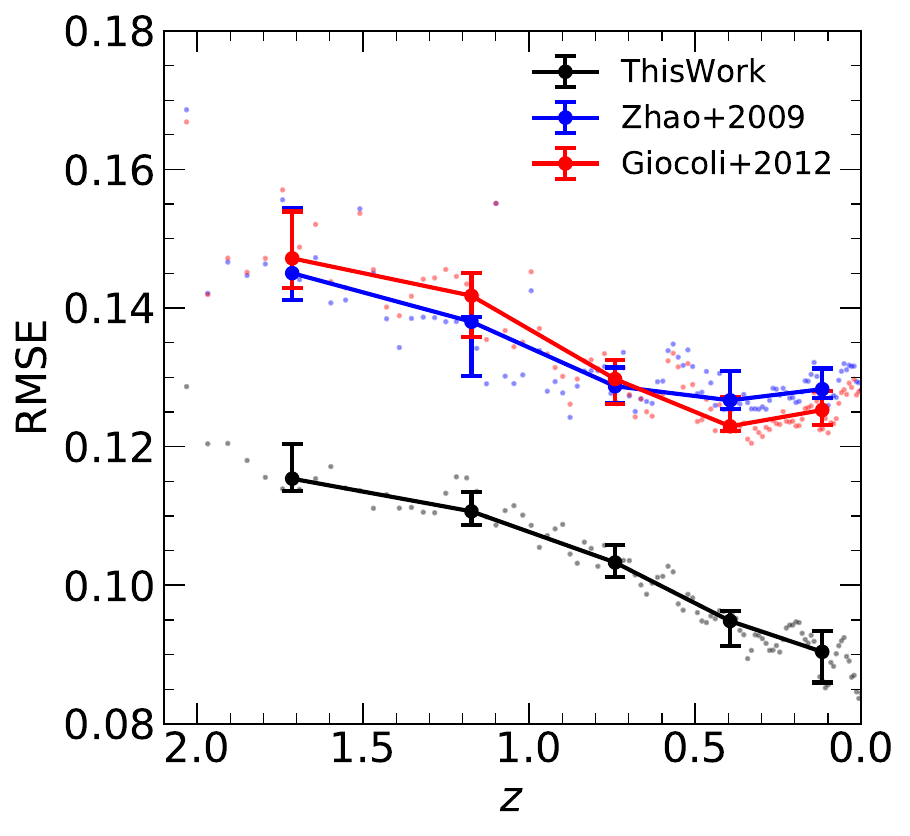} 
\caption{The median 
 relationship between the RMSE of the actual and predicted concentrations as a function of redshift. Black, blue, and red lines show results of neural network, Zhao, and Giocoli models, respectively. Error bars indicate the 16th and 84th percentiles. {The consistently lower RMSE of our model across redshifts suggests its superior predictive accuracy over the other models.}}\label{fig5}
\end{figure}

We further extend the prediction capability of our model to high-redshift samples, and present the evolution of the RMSE with redshift in Figure~\ref{fig5}. Both \citet{Zhao2009} and \citet{Giocoli2012} show similar RMSE values across different redshifts, with overall higher levels, indicating that these methods are less accurate in their predictions. Throughout the redshift range, the RMSE for our method remains consistently lower than that of the other two methods, decreasing as redshift decreases. {The change in RMSE from $z=0$ to $z=2$ is more obvious than that of the other two models. One possible reason is that as redshift increases, the model learns progressively less information about the mass accretion from halo merger trees, leading to a reduction in predictive capability.
In contrast, the other two models rely on fixed parameters, which results in their RMSE remaining relatively stable throughout the redshift range and consistently higher than that of our model.} Overall, our model demonstrates consistently lower RMSE across redshifts, indicating superior predictive accuracy compared to the other models.

\begin{figure}[H]
\includegraphics[width=0.8\linewidth]{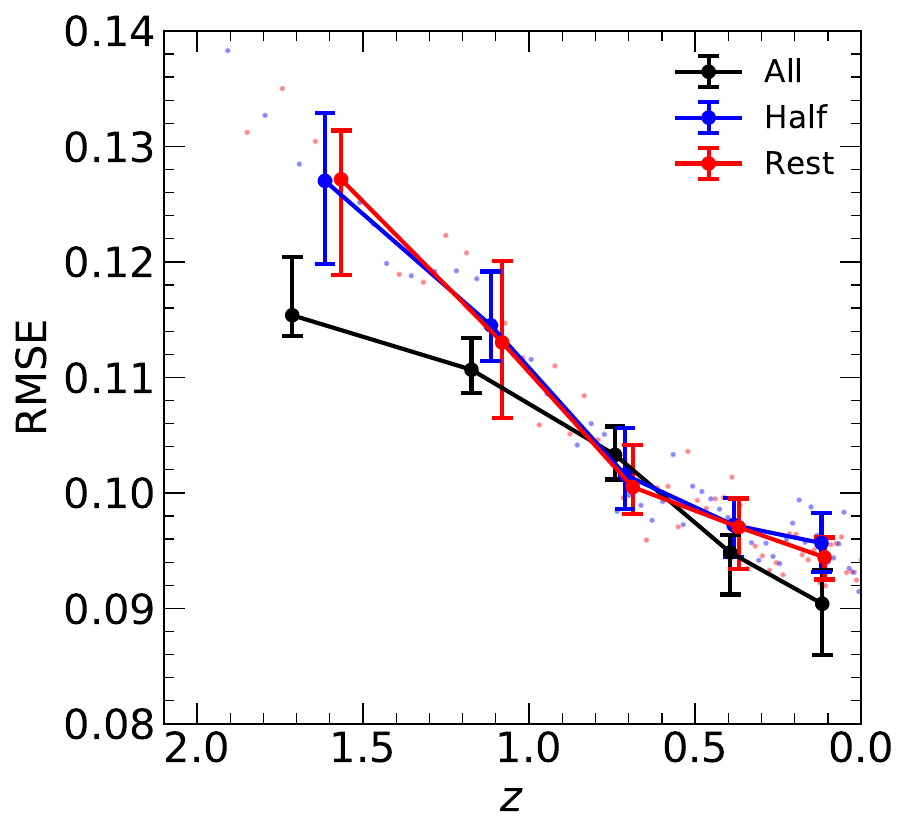} 
\caption{Similarly to 
 Figure~\ref{fig5}, the black line is identical to the one in the previous figure. The blue line represents the predictions on the TestDataset using the model trained on half of the snapshots between $z=2$ and $z=0$. Additionally, the red line shows the predictions made by the model from the blue line on the halo concentrations corresponding to the remaining half of the TestDataset snapshots. {This illustrates our model's ability to accurately predict halo concentrations for snapshots not present in the simulation.}}\label{fig6}
\end{figure}

We evaluate the model's ability to predict concentrations for snapshots not included in the simulation. Following the method outlined in Section \ref{sec_method_NN}, we train our model using half of the TrainDataset, sampled at equal intervals between $z=2$ and $z=0$. The trained model is then used to predict concentrations for half of the TestDataset snapshots within the same redshift range (blue line) and to track the RMSE evolution for the remaining TestDataset snapshots (red line) in Figure~\ref{fig6}. The close alignment of the two lines demonstrates that the model operates continuously rather than discretely. Given any halo mass accretion history and a target redshift, the model reliably predicts halo concentration at that redshift. Although halving the size of the TrainDataset slightly reduces model performance, it still outperforms the predictive capabilities of the models proposed by \citet{Zhao2009} and \citet{Giocoli2012}.

\section{Conclusions {and Outlook}}\label{sec_con}

Using the mass accretion histories of haloes from cosmological $N$-body simulations, we employ a deep learning neural network algorithm to predict the concentrations of individual dark haloes across various redshifts, achieving remarkable accuracy. Our key findings are summarized as follows:

\begin{itemize}

\item The neural network model accurately reproduces the established relationship between halo mass and halo concentration at $z=0$.

\item When tested on a new simulation with a different initial condition realization, the trained model performs exceptionally well. At $z=0$, the RMSE between the actual and predicted concentrations is approximately 0.08, which is significantly lower than the RMSE of about 0.13 obtained from the models of \citet{Zhao2009} and \citet{Giocoli2012}.

\item The model demonstrates robust predictive capability at other redshifts. Within the range $z=2$ to $z=0$, although the RMSE increases with redshift and prediction accuracy declines, the neural network consistently achieves substantially lower RMSE values compared to the models by \citet{Zhao2009} and \citet{Giocoli2012}.

\item The neural network model exhibits continuity in its predictions, enabling accurate estimation of halo concentrations for snapshots not explicitly included in the simulation.

\end{itemize}

Overall, our neural network model, trained on the mass accretion histories of haloes, demonstrates strong predictive power and robust performance. Given a merger tree and a target redshift, the model can reliably predict the concentration of individual dark matter haloes at that redshift.

{In this paper, our model fixes the cosmological parameters, leading to significant improvements in the corresponding concentration predictions. However, halo concentrations are also influenced by other factors. In a future work, we plan to explore models with varying cosmological parameters (such as $\Omega_{\rm m}$, $\Omega_{\Lambda}$ and $\sigma_8$) and investigate different halo definitions to further demonstrate the superiority, scalability, and robustness of our model.}

{Our current training set has some limitations, as the halo mass range it covers is relatively narrow. \citet{Wang2020} used cosmic zoom simulations to show that the mass-concentration relation, from the smallest Earth-mass haloes to the largest cluster-sized haloes, can be described by a single model~\citep{Zheng2024, Liu2024}. In the future, we plan to run a large number of high-resolution cosmological simulations of different box sizes to expand our training set to include a broader mass range to better compare with these studies} and explore the potential of neural networks to predict additional halo properties from their mass accretion histories.

\vspace{6pt} 

\authorcontributions{T.Z.: methodology and formal analysis; T.Z. and T.M.: software and data curation; T.Z.: writing---original draft; T.Z.: writing---review and editing; W.X. and G.L.: Visualization. All authors have read and agreed to the published version of the manuscript.}

\funding{This work was supported by the National Science Foundation of China(Grants Nos. 12403008, 62202446), Beijing Academy of Science and Technology Budding Talent Program(Grants No. 24CE-BGS-18), The Young Data Scientist Program of the China National Astronomical Data Center(Grants No. NADC2024YDS-03) and The Chongqing Natural Science Foundation(Grant No. cstc2021jcyj-msxmX0553).}

\dataavailability{The simulation data underlying this article will be shared on reasonable request to the corresponding author.} 

\acknowledgments{We thank Liang Gao, Jie Wang and Shihong Liao for useful discussions. Tianchi Zhang acknowledge support from the National Natural
Science Foundation of China (Grants No. 12403008), Beijing Academy of Science and Technology Budding Talent Program(Grants No. 24CE-BGS-18) and The Young Data Scientist Program of the China National Astronomical Data Center(Grants No. NADC2024YDS-03). Wenxiao Xu acknowledge support from the Chongqing Natural Science Foundation(Grant No. cstc2021jcyj-msxmX0553). Guan Li acknowledge support from the National Natural Science Foundation of China (Grants No. 62202446).}

\conflictsofinterest{The authors declare no conflicts of interest.} 

\begin{adjustwidth}{-\extralength}{0cm}

\reftitle{References}

\PublishersNote{}
\end{adjustwidth}
\end{document}